\documentclass[12pt]{article}
\usepackage[T1]{fontenc}
\usepackage[latin9]{inputenc}
\usepackage{amsmath}
\usepackage{graphicx}

\newcommand{\apj}{ApJ}

\newcommand{\mnras}{MNRAS}

\newcommand{\prd}{PRD}
\newcommand{\prl}{PRL}

\newcommand{\jmp}{J. Math. Phys.}

\usepackage[svgnames]{xcolor}
\usepackage[colorlinks=true,linkcolor=Blue,citecolor=Blue,urlcolor=Blue]{hyperref}

\title{Approximate Metric for a Rotating Deformed Mass}

\author{Francisco Frutos-Alfaro, 
$^{1,2}$\thanks{E-mail: frutos@fisica.ucr.ac.cr}, 
Paulo Montero-Camacho$^{1}$, \\ 
Miguel Araya-Arguedas$^{1,2}$, Javier Bonatti-Gonz\'alez$^{1,3}$ \\
$^{1}$School of Physics, University of Costa Rica, San Jos\'e, Costa Rica \\
$^{2}$Space Research Center, University of Costa Rica, San Jos\'e, Costa Rica \\
$^{3}$Nuclear Research Center, University of Costa Rica, San Jos\'e, Costa Rica}

\date{\today}

\begin{document}

\maketitle

\begin{abstract}
A new Kerr-like metric with quadrupole moment is obtained by means of 
perturbing the Kerr spacetime. The form of this new metric is simple as
the Kerr metric. By comparison with the exterior Hartle-Thorne 
metric, it is shown that it could be matched to an interior solution. 
This approximate metric may represent the spacetime of a real astrophysical 
object with any Kerr rotation parameter $ a $ and slightly deformed. 
\end{abstract}


\section{Introduction \label{sec:00}}

\noindent
In 1963, R.~P.~Kerr \cite{Kerr} proposed a metric that describes a massive 
rotating object. Since then, a huge amount of papers about the structure and 
astrophysical applications of this spacetime appeared. Now, it is widely 
believed that this metric does not represent the spacetime of an astrophysical 
rotating object. In 1967, Hern\'andez \cite{WHernandez} stated that 
{\it reasonable perfect fluid type solutions which might serve as source of 
the Kerr metric may not exist}. In 1971, Thorne \cite{Thorne,Marsh} added 
that {\it because of the relationship between multipole moments and angular 
momentum, the Kerr solution cannot represent correctly the external field of 
any realistic stars}. Moreover, the Kerr metric has difficulties when matching 
it to a realistic interior metric according to \cite{Boshkayev}. However, 
there has been a considerable amount of efforts trying to match the 
Kerr metric with a realistic interior metric that represents a physical source, 
see for example \cite{Cuchi,Glass,Haggag-Marek,Haggag2,Krasinski2,Ramadan}. 
For a concise and comprehensive review of the different methods that have been 
used in order to try and obtain an interior solution for the Kerr metric, 
see \cite{Krasinski}. 
 
\noindent
In \cite{Drake} and \cite{Viaggiu}, the Newman-Janis algorithm was applied to 
look for interior solutions. Drake and Turolla \cite{Drake} also propose a 
general method for finding interior solutions with oblate spheroidal boundary 
surfaces and note that the boundary surfaces reduce to a sphere in the case 
with no rotation, however Vaggiu in \cite{Viaggiu} argues that it is more 
helpful to start with the Schwarzschild interior and then proceed to the Kerr 
interior. Vaggiu uses an anisotropic conformally flat static interior and is 
lead to interior Kerr solutions with oblate spheroidal boundary surfaces, 
additionally he points out that his procedure can be applied to find interior 
solutions matching with a general asymptotically flat vacuum stationary 
spacetime.

\noindent
Other exact rotating solutions to the Einstein field equations (EFE) containing 
mass multipoles and magnetic dipole were obtained by 
\cite{Castejon,Manko,Manko2000,Pachon,Quevedo1986,Quevedo1989,Quevedo1991,
Quevedo2011}. 
In the first four articles, they used the Ernst formalism 
\cite{Ernst}, while in the four last ones, the solutions were obtained with 
the help of the Hoenselaers-Kinnersley-Xanthopoulos (HKX) transformations 
\cite{Hoenselaers}. These authors obtain new metrics from a given seed metric. 
These formalisms allow to include other desirable characteristics 
(rotation, multipole moments, magnetic dipole, etc.) to a given seed metrics. 
Furthermore, Quevedo in \cite{Quevedo2012} not only presents an exact 
electrovacuum solution that can be used to describe the exterior gravitational 
field of a rotating charged mass distribution, but also considers the 
matching using the derivatives of the curvature eigenvalues, this leads to 
matching conditions from which one can expect to obtain the minimum radius at 
which the matching can be made. 

\noindent
In Nature, it is expected that astrophysical objects are rotating and 
slightly deformed as is pointed out in \cite{Andersson} and in 
\cite{Stergioulas}. In addition, Andersson and Comer in \cite{Andersson} use 
a two fluid model for a neutron star one, layer with neutrons that has a 
differential rotation and another layer consisting of a solid crust with 
constant rotation. The aim of this article is to derive an appropriate 
analytical tractable metric for calculations in which the quadrupole moment 
can be treated as perturbation, but for arbitrary angular momentum. Moreover, 
this metric should be useful to tackle astrophysical problems, for instance, 
accretion disk in compact stellar objects 
\cite{Andersson,Fragile,Hawley,Stergioulas}, relativistic magnetohydrodynamic 
jet formation \cite{Fendt}, astrometry \cite{Frutos,Soffel} and gravitational 
lensing \cite{Frutos2001}. Furthermore, software related with 
applications of the Kerr metric can be easily modified in order to include the 
quadrupole moment \cite{Dexter,Frutos2012,Vincent}.

\noindent
This paper is organized as follows. In section 2, we give a succinct 
explanation of the Kerr metric, and the weak limit of the Erez-Rosen metric 
is presented. In section 3, the Lewis metric is presented, and the perturbation 
method is discussed. The application of this method leads to a new solution to 
the EFE with quadrupole moment and rotation. It is checked by means of the 
REDUCE software \cite{Hearn} that the resulting metric is a solution of 
the EFE. In section 4, we compare our solution with the exterior Hartle-Thorne 
metric in order to assure that our metric has astrophysical meaning. 
Forthcoming works with this metric are discussed inx section 6.

\section{The Kerr Metric and The Erez-Rosen Metric \label{sec:01}}

\subsection{The Kerr Metric \label{subsec:01}}

\noindent
The Kerr metric represents the spacetime of a non-deformed massive rotating 
object. The Kerr metric is given by \cite{Kerr,Carmeli}

\begin{eqnarray}
\label{kerr}
{d} {s}^{2} & = & \frac{\Delta}{{\rho}^2} [d t - a {\sin}^2 {\theta} d \phi]^2 
- \frac{{\sin}^2 {\theta}}{{\rho}^2} [(r^2 + a^2) d \phi - a d t]^2 \nonumber \\
& - & \frac{{\rho}^2}{\Delta} d {r}^2 - {\rho}^2 d {\theta}^2 ,  
\end{eqnarray}

\noindent
where $ \Delta = {r}^2 - 2 M r + {a}^2 $ and 
$ {\rho}^2 = {r}^2 + {a}^2 {\cos}^2 {\theta} $. $ M $ and $ a $ represent the 
mass and the rotation parameter, respectively. The angular momentum of the 
object is $ J = M a $.

\subsection{The Erez-Rosen Metric \label{subsec:02}}

\noindent 
The Erez-Rosen metric \cite{Carmeli,Winicour,Young,Zeldovich} 
represents the spacetime of a body with quadrupole moment. The principal axis 
of the quadrupole moment is chosen along the spin axis, so that gravitational 
radiation can be ignored. Here, we write down an approximate expression for 
this metric obtained by doing Taylor series \cite{Frutos}

\begin{eqnarray}
\label{erezrosen}
{d} {s}^{2} & = & \left({1-\frac{2M}{r}}\right) {\rm e}^{-2 \chi} {d} t^{2} 
- \left({1-\frac{2M}{r}}\right)^{-1} {\rm e}^{2 \chi} {{d} r^{2}} \nonumber \\ 
& - & {r^{2}}{\rm e}^{2 \chi} ({d} {\theta}^{2} + \sin^2{\theta} {d} {\phi}^{2}) ,
\end{eqnarray}

\noindent
where $ {d} {\Sigma}^2 = {d} {\theta}^{2} + \sin^2{\theta} {d} {\phi}^{2} $, 
and 

\begin{equation}
\label{chi}
\chi = \frac{2}{15} q \frac{M^{3}}{r^{3}} P_{2}(\cos{\theta}) ,
\end{equation}

\noindent
where $ P_2(\cos{\theta}) = {(3 \cos^2{\theta} - 1)}/{2} $. The quadrupole 
parameter is given by $ q = 15 G Q / (2 c^{2} M^{3}) $, with $ Q $ 
representing the quadrupole moment. This metric is valid up to the order 
$ O(q M^4, \, q^2) $.

\section{Perturbing the Kerr Metric \label{sec:02}}

\subsection{The Lewis Metrics \label{subsec:03}}

\noindent 
The Lewis metric is given by \cite{Lewis,Carmeli} 

\begin{equation}
\label{lewis} 
{d}{s}^2 = V d t^2 - 2 W d t d \phi 
- {\rm e}^{\mu} d \rho^2 - {\rm e}^{\nu} d z^2 - Z d \phi^2 ,
\end{equation}

\noindent 
where we have chosen the canonical co\-or\-di\-na\-tes $ x^{1} = \rho $ and 
$ x^{2} = z $, $ V, \, W, \, Z $, $ \mu $ and $ \nu $ are functions of 
$ \rho $ and $ z $ ($ \rho^2 = V Z + W^2 $). Choosing $ \mu = \nu $ and 
performing the following changes of potentials

$$ V = f , \quad W = \omega f , \quad Z = \frac{\rho^2}{f} - \omega^2 f 
\quad {\rm and} \quad {\rm e}^{\mu} = \frac{{\rm e}^{\gamma}}{f} , $$

\noindent 
we get the Papapetrou metric

\begin{equation}
\label{papapetrou} 
{d}{s}^2 = f (d t - \omega d \phi)^2 
- \frac{{\rm e}^{\gamma}}{f} [d \rho^2 + d z^2] - \frac{\rho^2}{f} d \phi^2 .
\end{equation}

\subsection{The Perturbation Method \label{subsec:04}}

\noindent
To include a small quadrupole moment into the Kerr metric we will modify the 
Lewis-Pa\-pa\-pe\-trou metric (\ref{papapetrou}). First of all, we choose 
expressions for the canonical coordinates $ \rho $ and $ z $. For the Kerr 
metric \cite{Kerr}, one particular choice is \cite{Carmeli,Chandrasekhar} 

\begin{equation}
\label{chandra} 
\rho = \sqrt{\Delta} \sin{\theta} \qquad {\rm and} \qquad 
z = (r - M) \cos{\theta} ,
\end{equation}

\noindent 
where $ \Delta = r^2 - 2 M r + a^2 $.

\noindent 
From (\ref{chandra}) we get

\begin{equation}
\label{cylindric}
d \rho^2 + d z^2 = 
[ {(r - M)^2} \sin^2{\theta} + \Delta \cos^2{\theta} ] 
\left( \frac{d r^2}{\Delta} + d {\theta}^2 \right) .
\end{equation}

\noindent 
If we choose

$$ {{\rm e}^{\mu}} 
= {\tilde{\rho}}^2 [ {(r - M)^2} \sin^2{\theta} 
+ \Delta \cos^2{\theta} ]^{-1} , $$

\noindent 
the term (\ref{cylindric}) becomes

$$ e^{\mu} [ d \rho^2 + d z^2 ] = {\tilde{\rho}}^2 
\left(\frac{{{d} r^{2}}}{\Delta} + {d} {\theta}^{2} \right) , $$

\noindent
where $ {\tilde{\rho}}^2 = r^2 + a^2 \cos^2{\theta} $.

\noindent 
From (\ref{papapetrou}), we propose the following metric 

\begin{equation}
\label{lewis2} 
{d}{s}^2 = {\cal V} d t^2 - 2 {\cal W} d t d \phi - {\cal X} {{d} r^{2}} 
- {\cal Y} {d} \theta^{2} - {\cal Z} {d} \phi^{2} , 
\end{equation}

\noindent 
where

\begin{eqnarray}
{\cal V} & = & V {\rm e}^{- 2 \psi} \nonumber \\
{\cal W} & = & W \nonumber \\
{\cal X} & = & X {\rm e}^{2 \psi} \\
{\cal Y} & = & Y {\rm e}^{2 \psi} \nonumber \\
{\cal Z} & = & Z {\rm e}^{2 \psi} , \nonumber 
\end{eqnarray}

\noindent 
where the potentials $ V, \, W, \, X, \, Y, \, Z $, and $ \psi $ depend on 
$ x^1 = r $ and $ x^2 = \theta $. The potential $ {\cal W} = W $ is so chosen 
to maintain the same cross components of the Kerr metric. 

\noindent 
Now, let us choose 

\begin{eqnarray}
V & = & f = \frac{1}{{\tilde{\rho}}^2} [\Delta - a^2 \sin^2{\theta}] 
\nonumber \\ 
W & = & \frac{a}{{\tilde{\rho}}^2} [\Delta - (r^2 + a^2)] \sin^2{\theta} 
= - \frac{2 J r}{{\tilde{\rho}}^2} \sin^2{\theta} \nonumber \\
X & = & \frac{{\tilde{\rho}}^2}{\Delta} \\
Y & = & {\tilde{\rho}}^2 \nonumber \\
Z & = & \frac{\sin^2{\theta}}{{\tilde{\rho}}^2} 
[(r^2 + a^2)^2 - a^2 \Delta \sin^2{\theta}] . \nonumber 
\end{eqnarray}

\noindent
The only potential we have to find is $ \psi $. In order to obtain this 
potential, the EFE must be solved 

\begin{equation}
\label{einstein} 
G_{i j} = R_{i j} - \frac{R}{2} g_{i j} = 0 ,
\end{equation}

\noindent 
where $ R_{i j} $ ($ i, \, j = 0, \, 1,\, 2, \, 3 $) are the Ricci tensor 
components and $ R $ is the curvature scalar. The Ricci tensor components and 
the curvature scalar $ R $ for this metric can be found in the Appendix. 

\noindent 
In our calculations, we consider the potential $ \psi $ as perturbation, 
{\it i.e.} one neglects terms of the form 

$$ \left(\frac{\partial \psi}{\partial r} \right)^2 = 
\left(\frac{\partial \psi}{\partial \theta} \right)^2 = 
\frac{\partial \psi}{\partial r} \frac{\partial \psi}{\partial \theta} 
\sim 0 . $$ 


\noindent 
Terms containing factors of the form 

$$ a \frac{\partial \psi}{\partial x^i} 
= m \frac{\partial \psi}{\partial x^i} \sim 0 
\qquad (i = 1, \, 2) $$

\noindent 
are also neglected. Substituting the known potentials 
($ V, \, W, \, X, \, Y, \, Z $) into the expressions for 
the Ricci tensor and the curvature scalar (see Appendix), it results only one 
equation for $ \psi $ that we have to solved:

\begin{equation}
\label{eqdif}
r^2 \sin{\theta} \, \nabla^2 \psi = \sin{\theta} 
\frac{\partial}{\partial r} 
\left(r^2 \frac{\partial \psi}{\partial r} \right) 
+ \frac{\partial}{\partial \theta} \left(\sin{\theta} 
\frac{\partial \psi}{\partial \theta} \right) = 0 .
\end{equation}

\noindent 
The solution for this equation is

\begin{equation}
\label{solution}
\psi = \frac{\cal K}{r^3} P_{2} (\cos{\theta}) , 
\end{equation}

\noindent 
where $ {\cal K} $ is a constant. To determine this constant, we compare the 
weak limit of the metric (\ref{lewis2}) with the approximate Erez-Rosen metric 
(\ref{erezrosen}). The result is $ {\cal K} = 2 q M^3 / 15 $ $ (\psi = \chi) $.

\noindent 
Then, the new modified Kerr metric containing quadrupole moment is

\begin{eqnarray}
\label{newkerr}
{d} {s}^{2} & = & \frac{{\rm e}^{- 2 \chi}}{{\rho}^2}
[\Delta - {a}^2 {\sin}^2 {\theta}] d {t}^2 
+ \frac{4 J r}{{\rho}^2} {\sin}^2 {\theta} d {t} d {\phi} \nonumber \\
& - & \frac{{\rho}^2 {\rm e}^{2 \chi}}{\Delta} d {r}^2 
- {\rho}^2 {\rm e}^{2 \chi} d {\theta}^2 \nonumber \\
& - & \frac{{\rm e}^{2 \chi} {\sin}^2 {\theta}}{{\rho}^2} 
[({r}^2 + {a}^2)^2 - {a}^2 \Delta {\sin}^2 {\theta}] d {\phi}^2 \nonumber \\
& = & \frac{\Delta}{{\rho}^2} [{\rm e}^{- \chi} d t 
- a {\rm e}^{\chi} {\sin}^2 {\theta} d \phi]^2 \nonumber \\
& - & \frac{{\sin}^2 {\theta}}{{\rho}^2} [(r^2 + a^2) {\rm e}^{\chi} d \phi 
- a {\rm e}^{- \chi} d t]^2 \nonumber \\
& - & {\rm e}^{2 \chi} \left(\frac{{\rho}^2}{\Delta} d {r}^2 
+ {\rho}^2 d {\theta}^2 \right) , 
\end{eqnarray}

\noindent 
where the tilde over $ \rho $ is dropped.

\noindent 
We verified that the metric \eqref{newkerr} is indeed a solution of the EFE 
using REDUCE \cite{Hearn} up to the order $ O(q M^4, \, q^2) $.

\noindent
Note that \eqref{newkerr} has four important limiting cases. One obtains the 
Kerr metric \eqref{kerr} if $ q = 0 $, the weak metric of \cite{Frutos} 
if $ a^2 = q^2 = q M^4 \simeq 0 $, the Erez-Rosen-like metric 
\eqref{erezrosen} if $ a = 0 $, and the Schwarzchild metric if $ q = a = 0 $.

\section{Comparison with the Exterior Hartle-Thorne Metric \label{sec:03}}

\noindent 
In order to validate the metric (\ref{newkerr}) as representing 
the gravitational field of a real astrophysical object, one should show that 
it is possible to construct an interior solution, which can appropriately be 
matched with our exterior solution. To this aim, we employed the exterior 
Hartle-Thorne metric \cite{Hartle,Berti,Boshkayev,Frutos}

\begin{eqnarray}
\label{hartle}
d s^2 & = & 
\left(1 - \frac{2 {\cal M}}{r} 
+ \frac{2 {\cal Q} {\cal M}^3}{r^3} P_2(\cos{\theta}) \right) d t^2 
\nonumber \\
& - & \left(1 + \frac{2 {\cal M}}{r} + \frac{4 {\cal M}^2}{r^2} 
- \frac{2 {\cal Q} {\cal M}^3 }{r^3} P_2(\cos{\theta}) \right) d r^2 \\
& - & r^2 \left(1 - \frac{2 {\cal Q} {\cal M}^3}{r^3} P_2(\cos{\theta}) 
\right) d \Sigma^2 + \frac{4 {\cal J}}{r} \sin^2{\theta} d t d \phi , 
\nonumber 
\end{eqnarray}

\noindent 
where $ {\cal M} $, $ {\cal J} $, and $ {\cal Q} $ are related with the total 
mass, angular momentum, and mass quadrupole moment of the rotating object, 
respectively. This approximation for the Hartle-Thorne metric \eqref{hartle} 
was obtained by Frutos-Alfaro {\it et al.} using a REDUCE program 
\cite{Frutos}. 

\noindent 
The spacetime (\ref{newkerr}) has the same weak limit as the metric obtained 
by Frutos-Alfaro {\it et al.} \cite{Frutos}. A comparison of the exterior 
Hartle-Thorne metric \eqref{hartle} with the weak limit of the metric 
(\ref{newkerr}) shows that upon defining 

\begin{equation}
\label{definitions}
{\cal M} = M, \qquad {\cal J} = J, \qquad 
2 {\cal Q} {\cal M}^3 = - \frac{4}{15} q {M^3} , 
\end{equation}

\noindent 
both metrics coincide up to the order $ O(M^3, \, a^2, \, q M^4, \, q^2) $. 
Hence, the metric (\ref{newkerr}) may be used to represent a compact 
astrophysical object.

\section{Conclusions \label{sec:04}}

\noindent
The new Kerr metric with quadrupole moment was obtained by solving the 
EFE approximately. It may represent the spacetime of a rotating and slightly 
deformed astrophysical object. This is possible, because it could be matched 
to an interior solution. We showed this by comparison of our metric with the 
exterior Hartle-Thorne metric. The limiting cases for the new Kerr metric 
correspond to the Kerr metric, the Erez-Rosen-like metric 
(see section \ref{sec:03}), and the Schwarzschild metric as expected. 

\noindent
The inclusion of the quadrupole moment in the Kerr metric does it more suitable 
for astrophysical calculations than the Kerr metric alone. There are a large 
variety of applications which can be tackled with this new metric. 
Amongst the applications for this metric are astrometry, gravitational lensing, 
relativistic magnetohydrodynamic jet formation, and accretion disks in compact 
stellar objects, additionally we would like to point out that works in 
superfluid neutron stars can be repeated but using this new metric instead of 
the Hartle-Thorne metric as an exterior solution. Furthermore, the existing 
software with applications of the Kerr metric can be easily modified to include 
the quadrupole moment.


\appendix
\section{Appendix}

\noindent
The non-null Ricci tensor components for the metric \eqref{lewis2} are given by 
(with the tilde over $ \rho $ dropped)

\begin{eqnarray*}
R_{0 0} & = & \frac{{\rm e}^{- 2 \psi}}{4 \rho^2 X^2 Y^2} \left( 
- 4 \rho^2 V X^2 Y \frac{\partial^2 \psi}{\partial \theta^2} \right. \\
& + & \left. 8 V W^2 X^2 Y 
\left(\frac{\partial \psi}{\partial \theta} \right)^2 
- 2 \rho^2 V X Y \frac{\partial \psi}{\partial \theta} 
\frac{\partial X}{\partial \theta} \right. \\
& + & \left. 2 V X^2 Y \frac{\partial \psi}{\partial \theta} 
\frac{\partial \rho^2}{\partial \theta} 
- 4 \rho^2 X^2 Y \frac{\partial \psi}{\partial \theta} 
\frac{\partial V}{\partial \theta} \right. \\
& - & \left. 4 W^2 X^2 Y \frac{\partial \psi}{\partial \theta} 
\frac{\partial V}{\partial \theta} 
+ 2 \rho^2 V X^2  \frac{\partial \psi}{\partial \theta} 
\frac{\partial Y}{\partial \theta} \right. \\
& - & \left. 4 V^2 X^2 Y \frac{\partial \psi}{\partial \theta} 
\frac{\partial Z}{\partial \theta}
- 4 \rho^2 V X Y^2 \frac{\partial^2 \psi}{\partial r^2} \right. \\
& + & \left. 8 V W^2 X Y^2 \left(\frac{\partial \psi}{\partial r} \right)^2 
+ 2 \rho^2 V Y^2 \frac{\partial \psi}{\partial r} 
\frac{\partial X}{\partial r} \right. \\
& + & \left. 2 V X Y^2 \frac{\partial \psi}{\partial r} 
\frac{\partial \rho^2}{\partial r} 
- 4 \rho^2 X Y^2 \frac{\partial \psi}{\partial r} 
\frac{\partial V}{\partial r} \right. \\
& - & \left. 4 W^2 X Y^2 \frac{\partial \psi}{\partial r} 
\frac{\partial V}{\partial r} 
- 2 \rho^2 V X Y \frac{\partial \psi}{\partial r} 
\frac{\partial Y}{\partial r} \right. \\
& - & \left. 4 V^2 X Y^2 \frac{\partial \psi}{\partial r} 
\frac{\partial Z}{\partial r} 
+ \rho^2 X Y \frac{\partial X}{\partial \theta} 
\frac{\partial V}{\partial \theta} \right. \\
& - & \left. \rho^2 Y^2 \frac{\partial X}{\partial r} 
\frac{\partial V}{\partial r} 
- X^2 Y \frac{\partial \rho^2}{\partial \theta} 
\frac{\partial V}{\partial \theta} \right. \\
& - & \left. X Y^2 \frac{\partial \rho^2}{\partial r} 
\frac{\partial V}{\partial r} 
+ 2 \rho^2 X^2 Y \frac{\partial^2 V}{\partial \theta^2} \right. \\
& - & \left. \rho^2 X^2 \frac{\partial V}{\partial \theta} 
\frac{\partial Y}{\partial \theta} 
+ 2 V X^2 Y \frac{\partial V}{\partial \theta} 
\frac{\partial Z}{\partial \theta} \right. \\
& + & \left. 2 \rho^2 X Y^2 \frac{\partial^2 V}{\partial r^2} 
+ \rho^2 X Y \frac{\partial V}{\partial r} 
\frac{\partial Y}{\partial r} \right. \\
& + & \left. 2 V X Y^2 \frac{\partial V}{\partial r} 
\frac{\partial Z}{\partial r} 
+ 2 V X^2 Y \left(\frac{\partial W}{\partial \theta} \right)^2 \right. \\
& + & \left. 2 V X Y^2 \left(\frac{\partial W}{\partial r} \right)^2 \right) 
\end{eqnarray*}


\begin{eqnarray*}
R_{0 3} & = & \frac{{\rm e}^{- 2 \psi}}{4 \rho^2 X^2 Y^2} \left(
8 \rho^2 W X^2 Y \left(\frac{\partial \psi}{\partial \theta} \right)^2 
\right. \\ 
& - & \left. 8 W^3 X^2 Y \left(\frac{\partial \psi}{\partial \theta} \right)^2 
- 4 W X^2 Y \frac{\partial \psi}{\partial \theta} 
\frac{\partial \rho^2}{\partial \theta} \right. \\
& + & \left. 8 W^2 X^2 Y \frac{\partial \psi}{\partial \theta} 
\frac{\partial W}{\partial \theta} 
+ 8 V W X^2 Y \frac{\partial \psi}{\partial \theta} 
\frac{\partial Z}{\partial \theta} \right. \\
& + & \left. 8 \rho^2 W X Y^2 \left(\frac{\partial \psi}{\partial r} \right)^2 
- 8 W^3 X Y^2 \left(\frac{\partial \psi}{\partial r} \right)^2 \right. \\
& - & \left. 4 W X Y^2 \frac{\partial \psi}{\partial r} 
\frac{\partial \rho^2}{\partial r} 
+ 8 W^2 X Y^2 \frac{\partial \psi}{\partial r} \frac{\partial W}{\partial r} 
\right. \\
& + & \left. 8 V W X Y^2 \frac{\partial \psi}{\partial r} 
\frac{\partial Z}{\partial r} 
- \rho^2 X Y \frac{\partial X}{\partial \theta} 
\frac{\partial W}{\partial \theta} \right. \\
& + & \left. \rho^2 Y^2 \frac{\partial X}{\partial r} 
\frac{\partial W}{\partial r} 
+ X^2 Y \frac{\partial \rho^2}{\partial \theta} 
\frac{\partial W}{\partial \theta} \right. \\ 
& + & \left. X Y^2 \frac{\partial \rho^2}{\partial r} 
\frac{\partial W}{\partial r} 
- 2 W X^2 Y \frac{\partial V}{\partial \theta} 
\frac{\partial Z}{\partial \theta} \right. \\
& - & \left. 2 W X Y^2 \frac{\partial V}{\partial r} 
\frac{\partial Z}{\partial r} 
- 2 \rho^2 X^2 Y \frac{\partial^2 W}{\partial \theta^2} \right. \\
& - & \left. 2 W X^2 Y \left(\frac{\partial W}{\partial \theta} \right)^2 
+ \rho^2 X^2 \frac{\partial W}{\partial \theta} 
\frac{\partial Y}{\partial \theta} \right. \\
& - & \left. 2 \rho^2 X Y^2 \frac{\partial^2 W}{\partial r^2} 
- 2 W X Y^2 \left(\frac{\partial W}{\partial r} \right)^2 \right. \\
& - & \left. \rho^2 X Y \frac{\partial W}{\partial r} 
\frac{\partial Y}{\partial r} \right) = R_{3 0}
\end{eqnarray*}

\begin{eqnarray*}
R_{1 1} & = & \frac{1}{4 \rho^4 X Y^2} \left( 
- 4 \rho^4 X^2 Y \frac{\partial^2 \psi}{\partial \theta^2} 
- 2 \rho^4 X Y \frac{\partial \psi}{\partial \theta} 
\frac{\partial X}{\partial \theta} \right. \\
& - & \left. 2 \rho^2 X^2 Y \frac{\partial \psi}{\partial \theta} 
\frac{\partial \rho^2}{\partial \theta} 
+ 2 \rho^4 X^2 \frac{\partial \psi}{\partial \theta} 
\frac{\partial Y}{\partial \theta} \right. \\
& - & \left. 4 \rho^4 X Y^2 \frac{\partial^2 \psi}{\partial r^2} 
- 8 \rho^4 X Y^2 \left(\frac{\partial \psi}{\partial r} \right)^2 \right. \\
& + & \left. 8 \rho^2 W^2 X Y^2 
\left(\frac{\partial \psi}{\partial r} \right)^2 
+ 2 \rho^4 Y^2 \frac{\partial \psi}{\partial r} \frac{\partial X}{\partial r} 
\right. \\
& + & \left. 6 \rho^2 X Y^2 \frac{\partial \psi}{\partial r} 
\frac{\partial \rho^2}{\partial r} 
- 8 \rho^2 W X Y^2 \frac{\partial \psi}{\partial r} 
\frac{\partial W}{\partial r} \right. \\
& - & \left. 2 \rho^4 X Y \frac{\partial \psi}{\partial r} 
\frac{\partial Y}{\partial r} 
- 8 \rho^2 V X Y^2 \frac{\partial \psi}{\partial r} 
\frac{\partial Z}{\partial r} \right. \\
& - & \left. 2 \rho^4 X Y \frac{\partial^2 X}{\partial \theta^2} 
+ \rho^4 Y \left(\frac{\partial X}{\partial \theta} \right)^2 \right. \\
& - & \left. \rho^2 X Y \frac{\partial X}{\partial \theta} 
\frac{\partial \rho^2}{\partial \theta} 
+ \rho^4 X \frac{\partial X}{\partial \theta} 
\frac{\partial Y}{\partial \theta} \right. \\
& + & \left. \rho^2 Y^2 \frac{\partial X}{\partial r} 
\frac{\partial \rho^2}{\partial r} 
+ \rho^4 Y \frac{\partial X}{\partial r} \frac{\partial Y}{\partial r} 
\right. \\
& - & \left. 2 \rho^2 X Y^2 \frac{\partial^2 \rho^2}{\partial r^2} 
+ X Y^2 \left(\frac{\partial \rho^2}{\partial r} \right)^2 \right. \\
& + & \left. 2 V X Y^2 \frac{\partial \rho^2}{\partial r} 
\frac{\partial Z}{\partial r} 
+ 2 W^2 X Y^2 \frac{\partial V}{\partial r} \frac{\partial Z}{\partial r} 
\right. \\
& + & \left. 2 \rho^2 X Y^2 \left(\frac{\partial W}{\partial r} \right)^2 
- 4 V W X Y^2 \frac{\partial W}{\partial r} \frac{\partial Z}{\partial r} 
\right. \\
& - & \left. 2 \rho^4 X Y \frac{\partial^2 Y}{\partial r^2} 
+ \rho^4 X \left(\frac{\partial Y}{\partial r} \right)^2 \right. \\
& - & \left. 2 V^2 X Y^2 \left(\frac{\partial Z}{\partial r} \right)^2 \right)
\end{eqnarray*}


\begin{eqnarray*}
R_{1 2} & = & \frac{1}{4 \rho^4 X Y} \left( 
- 8 \rho^4 X Y \frac{\partial \psi}{\partial \theta} 
\frac{\partial \psi}{\partial r} \right. \\
& + & \left. 8 \rho^2 W^2 X Y \frac{\partial \psi}{\partial \theta} 
\frac{\partial \psi}{\partial r} 
+ 4 \rho^2 X Y \frac{\partial \psi}{\partial \theta} 
\frac{\partial \rho^2}{\partial r} \right. \\
& - & \left. 4 \rho^2 W X Y \frac{\partial \psi}{\partial \theta} 
\frac{\partial W}{\partial r} 
- 4 \rho^2 V X Y \frac{\partial \psi}{\partial \theta} 
\frac{\partial Z}{\partial r} \right. \\
& + & \left. 4 \rho^2 X Y \frac{\partial \psi}{\partial r} 
\frac{\partial \rho^2}{\partial \theta} 
- 4 \rho^2 W X Y \frac{\partial \psi}{\partial r} 
\frac{\partial W}{\partial \theta} \right. \\
& - & \left. 4 \rho^2 V X Y \frac{\partial \psi}{\partial r} 
\frac{\partial Z}{\partial \theta} 
+ \rho^2 Y \frac{\partial X}{\partial \theta} 
\frac{\partial \rho^2}{\partial r} \right. \\ 
& - & \left. 2 \rho^2 X Y \frac{\partial^2 \rho^2}{\partial \theta \partial r} 
+ W^2 X Y \frac{\partial^2 \rho^2}{\partial \theta \partial r} \right. \\
& + & \left. X Y \frac{\partial \rho^2}{\partial \theta} 
\frac{\partial \rho^2}{\partial r} 
+ \rho^2 X \frac{\partial \rho^2}{\partial \theta} 
\frac{\partial Y}{\partial r} \right. \\ 
& + & \left. V X Y \frac{\partial \rho^2}{\partial \theta} 
\frac{\partial Z}{\partial r} 
+ V X Y \frac{\partial \rho^2}{\partial r} \frac{\partial Z}{\partial \theta} 
\right. \\
& - & \left. W^2 X Y Z \frac{\partial^2 V}{\partial \theta \partial r} 
- 2 W^3 X Y \frac{\partial^2 W}{\partial \theta \partial r} \right. \\
& + & \left. 2 \rho^2 X Y \frac{\partial W}{\partial \theta} 
\frac{\partial W}{\partial r} 
- 2 W^2 X Y \frac{\partial W}{\partial \theta} \frac{\partial W}{\partial r} 
\right. \\
& - & \left. 2 V W X Y \frac{\partial W}{\partial \theta} 
\frac{\partial Z}{\partial r} 
-2 V W X Y \frac{\partial W}{\partial r} \frac{\partial Z}{\partial \theta} 
\right. \\
& - & \left. V W^2 X Y \frac{\partial^2 Z}{\partial \theta \partial r} 
- 2 V^2 X Y \frac{\partial Z}{\partial \theta} \frac{\partial Z}{\partial r} 
\right) = R_{2 1}
\end{eqnarray*}

\begin{eqnarray*}
R_{2 2} & = & \frac{1}{4 \rho^4 X^2 Y} \left( 
- 4 \rho^4 X^2 Y \frac{\partial^2 \psi}{\partial \theta^2} \right. \\
& - & \left. 8 \rho^4 X^2 Y 
\left(\frac{\partial \psi}{\partial \theta} \right)^2 
+ 8 \rho^2 W^2 X^2 Y \left(\frac{\partial \psi}{\partial \theta} \right)^2 
\right. \\
& - & \left. 2 \rho^4 X Y \frac{\partial \psi}{\partial \theta} 
\frac{\partial X}{\partial \theta} 
+ 6 \rho^2 X^2 Y \frac{\partial \psi}{\partial \theta} 
\frac{\partial \rho^2}{\partial \theta} \right. \\
& - & \left. 8 \rho^2 W X^2 Y \frac{\partial \psi}{\partial \theta} 
\frac{\partial W}{\partial \theta} 
+ 2 \rho^4 X^2 \frac{\partial \psi}{\partial \theta} 
\frac{\partial Y}{\partial \theta} \right. \\
& - & \left. 8 \rho^2 V X^2 Y \frac{\partial \psi}{\partial \theta} 
\frac{\partial Z}{\partial \theta} 
- 4 \rho^4 X Y^2 \frac{\partial^2 \psi}{\partial r^2} \right. \\
& + & \left. 2 \rho^4 Y^2 \frac{\partial \psi}{\partial r} 
\frac{\partial X}{\partial r} 
- 2 \rho^2 X Y^2 \frac{\partial \psi}{\partial r} 
\frac{\partial \rho^2}{\partial r} \right. \\
& - & \left. 2 \rho^4 X Y \frac{\partial \psi}{\partial r} 
\frac{\partial Y}{\partial r} 
- 2 \rho^4 X Y \frac{\partial^2 X}{\partial \theta^2} \right. \\ 
& + & \left. \rho^4 Y \left(\frac{\partial X}{\partial \theta} \right)^2 
+ \rho^4 X \frac{\partial X}{\partial \theta} 
\frac{\partial Y}{\partial \theta} \right. \\
& + & \left. \rho^4 Y \frac{\partial X}{\partial r} 
\frac{\partial Y}{\partial r} 
- 2 \rho^2 X^2 Y \frac{\partial^2 \rho^2}{\partial \theta^2} \right. \\
& + & \left. X^2 Y \left(\frac{\partial \rho^2}{\partial \theta} \right)^2 
+ \rho^2 X^2 \frac{\partial \rho^2}{\partial \theta} 
\frac{\partial Y}{\partial \theta} \right. \\
& + & \left. 2 V X^2 Y \frac{\partial \rho^2}{\partial \theta} 
\frac{\partial Z}{\partial \theta} 
- \rho^2 X Y \frac{\partial \rho^2}{\partial r} 
\frac{\partial Y}{\partial r} \right. \\
& + & \left. 2 W^2 X^2 Y \frac{\partial V}{\partial \theta} 
\frac{\partial Z}{\partial \theta} 
+ 2 \rho^2 X^2 Y \left(\frac{\partial W}{\partial \theta} \right)^2 \right. \\
& - & \left. 4 V W X^2 Y \frac{\partial W}{\partial \theta} 
\frac{\partial Z}{\partial \theta} 
- 2 \rho^4 X Y \frac{\partial^2 Y}{\partial r^2} \right. \\
& + & \left. \rho^4 X \left(\frac{\partial Y}{\partial r} \right)^2 
- 2 V^2 X^2 Y \left(\frac{\partial Z}{\partial \theta} \right)^2 \right)  
\end{eqnarray*}

\begin{eqnarray*}
R_{3 3} & = & \frac{1}{4 \rho^2 X^2 Y^2} \left( 
- 4 \rho^2 X^2 Y Z \frac{\partial^2 \psi}{\partial \theta^2} \right. \\
& - & \left. 8 W^2 X^2 Y Z 
\left(\frac{\partial \psi}{\partial \theta} \right)^2 
- 2 \rho^2 X Y Z \frac{\partial \psi}{\partial \theta} 
\frac{\partial X}{\partial \theta} \right. \\
& - & \left. 2 Y X^2 Z \frac{\partial \psi}{\partial \theta} 
\frac{\partial \rho^2}{\partial \theta} 
+ 8 W X^2 Y Z \frac{\partial \psi}{\partial \theta} 
\frac{\partial W}{\partial \theta} \right. \\ 
& + & \left. 2 \rho^2 X^2 Z \frac{\partial \psi}{\partial \theta} 
\frac{\partial Y}{\partial \theta} 
- 8 W^2 X^2 Y \frac{\partial \psi}{\partial \theta} 
\frac{\partial Z}{\partial \theta} \right. \\
& - & \left. 4 \rho^2 X Y^2 Z \frac{\partial^2 \psi}{\partial r^2} 
- 8 W^2 X Y^2 Z \left(\frac{\partial \psi}{\partial r} \right)^2 \right. \\
& + & \left. 2 \rho^2 Y^2 Z \frac{\partial \psi}{\partial r} 
\frac{\partial X}{\partial r} 
- 2 X Y^2 Z \frac{\partial \psi}{\partial r} 
\frac{\partial \rho^2}{\partial r} \right. \\ 
& + & \left. 8 W X Y^2 Z \frac{\partial \psi}{\partial r} 
\frac{\partial W}{\partial r} 
- 2 \rho^2 X Y Z \frac{\partial \psi}{\partial r} 
\frac{\partial Y}{\partial r} \right. \\
& - & \left. 8 W^2 X Y^2 \frac{\partial \psi}{\partial r} 
\frac{\partial Z}{\partial r} 
- \rho^2 X Y \frac{\partial X}{\partial \theta} 
\frac{\partial Z}{\partial \theta} \right. \\
& + & \left. \rho^2 Y^2 \frac{\partial X}{\partial r} 
\frac{\partial Z}{\partial r} 
- X^2 Y \frac{\partial \rho^2}{\partial \theta} 
\frac{\partial Z}{\partial \theta} \right. \\ 
& - & \left. X Y^2 \frac{\partial \rho^2}{\partial r} 
\frac{\partial Z}{\partial r} 
- 2 X^2 Y Z \left(\frac{\partial W}{\partial \theta} \right)^2 \right. \\
& + & \left. 4 W X^2 Y \frac{\partial W}{\partial \theta} 
\frac{\partial Z}{\partial \theta} 
- 2 X Y^2 Z \left(\frac{\partial W}{\partial r} \right)^2 \right. \\
& + & \left. 4 W X Y^2 \frac{\partial W}{\partial r} 
\frac{\partial Z}{\partial r} 
+ \rho^2 X^2 \frac{\partial Y}{\partial \theta} 
\frac{\partial Z}{\partial \theta} \right. \\ 
& - & \left. \rho^2 X Y \frac{\partial Y}{\partial r} 
\frac{\partial Z}{\partial r} 
- 2 \rho^2 X^2 Y \frac{\partial^2 Z}{\partial \theta^2} \right. \\
& + & \left. 2 V X^2 Y \left(\frac{\partial Z}{\partial \theta} \right)^2 
- 2 \rho^2 X Y^2 \frac{\partial^2 Z}{\partial r^2} \right. \\ 
& + & \left. 2 V X Y^2 \left(\frac{\partial Z}{\partial r} \right)^2 \right)
\end{eqnarray*}


\noindent
Calculation of the scalar curvature (with the tilde over $ \rho $ dropped)

\begin{eqnarray*}
R & = & \frac{{\rm e}^{- 2 \psi} }{2 \rho^4 X^2 Y^2} \left(
4 \rho^4 X^2 Y \frac{\partial^2 \psi}{\partial \theta^2} 
+ 4 \rho^4 X^2 Y \left(\frac{\partial \psi}{\partial \theta} \right)^2 
\right. \\ 
& - & \left. 4 \rho^2 W^2 X^2 Y \left(\frac{\partial \psi}{\partial \theta} 
\right)^2 
+ 2 \rho^4 X Y \frac{\partial \psi}{\partial \theta} 
\frac{\partial X}{\partial \theta} \right. \\
& - & \left. 2 \rho^2 X^2 Y \frac{\partial \psi}{\partial \theta} 
\frac{\partial \rho^2}{\partial \theta} 
+ 4 \rho^2 W X^2 Y \frac{\partial \psi}{\partial \theta} 
\frac{\partial W}{\partial \theta} \right. \\
& - & \left. 2 \rho^4 X^2 \frac{\partial \psi}{\partial \theta} 
\frac{\partial Y}{\partial \theta} 
+ 4 \rho^2 V X^2 Y \frac{\partial \psi}{\partial \theta} 
\frac{\partial Z}{\partial \theta} \right. \\ 
& + & \left. 4 \rho^4 X Y^2 \frac{\partial^2 \psi}{\partial r^2} 
+ 4 \rho^4 X Y^2 \left(\frac{\partial \psi}{\partial r} \right)^2 \right. \\
& - & \left. 4 \rho^2 W^2 X Y^2 
\left(\frac{\partial \psi}{\partial r} \right)^2 
- 2 \rho^4 Y^2 \frac{\partial \psi}{\partial r} \frac{\partial X}{\partial r} 
\right. \\
& - & \left. 2 \rho^2 X Y^2 \frac{\partial \psi}{\partial r} 
\frac{\partial \rho^2}{\partial r} 
+ 4 \rho^2 W X Y^2 \frac{\partial \psi}{\partial r} 
\frac{\partial W}{\partial r} \right. \\ 
& + & \left. 2 \rho^4 X Y \frac{\partial \psi}{\partial r} 
\frac{\partial Y}{\partial r} 
+ 4 \rho^2 V X Y^2 \frac{\partial \psi}{\partial r} 
\frac{\partial Z}{\partial r} \right. \\ 
& + & \left. 2 \rho^4 X Y \frac{\partial^2 X}{\partial \theta^2} 
- \rho^4 Y \left(\frac{\partial X}{\partial \theta} \right)^2 \right. \\
& + & \left. \rho^2 X Y \frac{\partial X}{\partial \theta} 
\frac{\partial \rho^2}{\partial \theta} 
- \rho^4 X \frac{\partial X}{\partial \theta} 
\frac{\partial Y}{\partial \theta} \right. \\ 
& - & \left. \rho^2 Y^2 \frac{\partial X}{\partial r} 
\frac{\partial \rho^2}{\partial r} 
- \rho^4 Y \frac{\partial X}{\partial r} \frac{\partial Y}{\partial r} 
\right. \\
& + & \left. 2 \rho^2 X^2 Y \frac{\partial^2 \rho^2}{\partial \theta^2} 
- X^2 Y \left(\frac{\partial \rho^2}{\partial \theta} \right)^2 \right. \\
& - & \left. \rho^2 X^2 \frac{\partial \rho^2}{\partial \theta} 
\frac{\partial Y}{\partial \theta} 
+ 2 \rho^2 X Y^2 \frac{\partial^2 \rho^2}{\partial r^2} \right. \\ 
& - & \left. X Y^2 \left(\frac{\partial \rho^2}{\partial r} \right)^2 
+ \rho^2 X Y \frac{\partial \rho^2}{\partial r} \frac{\partial Y}{\partial r} 
\right. \\
& - & \left. \rho^2 X^2 Y \frac{\partial V}{\partial \theta} 
\frac{\partial Z}{\partial \theta} 
- \rho^2 X Y^2 \frac{\partial V}{\partial r} \frac{\partial Z}{\partial r} 
\right. \\
& - & \left. \rho^2 X^2 Y \left(\frac{\partial W}{\partial \theta} \right)^2 
- \rho^2 X Y^2 \left(\frac{\partial W}{\partial r} \right)^2 \right. \\ 
& + & \left. 2 \rho^4 X Y \frac{\partial^2 Y}{\partial r^2} 
- \rho^4 X \left(\frac{\partial Y}{\partial r} \right)^2 \right)
\end{eqnarray*}


\end{document}